\documentclass[%
 reprint,
 amsmath,amssymb,
prfluids,
floatfix,
]{revtex4-2}
\usepackage{amsmath}
\usepackage{bm}
\usepackage{graphicx}
\usepackage{ascmac}
\usepackage{mathtools}
\usepackage{color}
\usepackage{overpic}

\begin{document}

\preprint{APS/123-QED}
\title{
Phase-symmetry breaking as a mechanism for subcritical transition
in shell models of turbulence
}

\author{Yoshiki Hiruta}
 \email{hiruta@rs.tus.ac.jp}
  \affiliation{
	Department of Physics and Astronomy, Faculty of Science and Technology, Tokyo University of Science, Chiba 278-8510, Japan}
\begin{abstract}

Subcritical transition to turbulence, in which the
laminar state is linearly stable yet finite-amplitude
perturbations develop into turbulence, is ubiquitous
but lacks a simple analytical framework.
We demonstrate such a framework using a shell model of
turbulence, in which external forcing breaks the phase
symmetry of the governing equations. This symmetry
breaking suppresses the linear instability of the laminar
state, 
while the energy
cascade and spectrum of the developed turbulent
state are preserved. A complementary single-triad model
admits an exact elliptic neutral stability curve, revealing
that the stabilization depends only on the breaking
strength and not on the nonlinear coupling coefficients.
Since the phase symmetry of the shell model corresponds
to Galilean invariance in the Navier--Stokes equations,
this mechanism may offer a new perspective on subcritical
transition in fluid systems.

\end{abstract}

\maketitle

\textit{Introduction.}---
Turbulence often arises subcritically: small disturbances
decay while sufficiently large ones develop into turbulence,
even in the absence of laminar-state instability
~\cite{itano_2001,kawahara_2001,willis2007critical,shimizu2009driving,kawahara_2012,linkmann2015sudden,barkley2015rise,barkley2016theoretical,veen2016sub,hiruta_2020,hiruta2022,avila2023transition,hiruta2025bc}.
Understanding the sustenance of turbulence and identifying
the minimal perturbation to trigger it remain central
problems, because subcritical transitions are accompanied by
enhanced energy dissipation~\cite{Pringle2010,cherubini2011minimal,Pringle2012,chu2024minimal},
large-scale laminar--turbulent patterns~\cite{barkley2005computational,hiruta_2015,hiruta_2017,matsukawa2025switching,mukund2025aging,manneville2025critical},
and deep connections to nonequilibrium phase
transitions~\cite{lemoult2016directed,sano2016universal,shimizu2019,hiruta_2020,hiruta2022,kohyama2022sidewall,hof2023directed}.
Owing to the linear stability of the base state, however, a fully nonlinear treatment is required even at the onset of subcritical transition,
since linear and weakly nonlinear analyses can only describe the decay dynamics toward the stable laminar state.

Unlike supercritical transitions, subcritical transitions are
difficult to predict without extensive computation. Existing
approaches based on bifurcation and linear stability
analyses~\cite{hamilton1995regeneration,waleffe1995,waleffe1997,avila2013,itano_2001,kawahara_2001,gibson2008visualizing,kawahara_2012,deguchi2013emergence,linkmann2015sudden,graham2021exact,page2024exact,weyrauch2025chaotic}
are highly
system-dependent and lack simple analytical frameworks,
leaving reliable indicators for subcritical transitions scarce.

To make analytical progress, we focus on symmetry breaking
of the governing equations without specifying a particular
flow-driving condition, following the insight that
low-dimensional models can isolate key mechanisms of
subcritical transition~\cite{hamilton1995regeneration,waleffe1995,waleffe1997,barkley2016theoretical}. Symmetry
breaking plays a determining role across diverse fields,
from dark matter to active matter~\cite{marchetti2013hydrodynamics,delacretaz2022damping,chala2022review}.
The shell model of turbulence~\cite{yamada1987lyapunov,l1998improved,kadanoff1995scaling,biferale2003shell,kato2024laminar,tanogami2024information}, a
simplified model retaining the essential energy cascade,
possesses a phase symmetry associated with the Galilean
invariance of the Navier--Stokes
equations~\cite{bernard2007inverse,oberlack2001unified,oberlack2022turbulence,mailybaev2022hidden}. This symmetry characterizes
nonlinear behaviors such as energy transfer and scaling
laws~\cite{ditlevsen2000symmetries,mailybaev2021hidden,mailybaev2023hidden,verma2023renormalization,mailybaev2025rg,benavides2025phase,manfredini2026nonlinear}.

We investigate the role of phase-symmetry breaking induced
by external forcing in subcritical transition within the
shell model. In fully developed turbulence, the
high-dimensional chaotic dynamics are not constrained by a
single symmetry, yet the same symmetry can strongly
restrict stability.
To exploit this contrast, we complement the shell
model with a single-triad (ST) model that retains the
essential symmetry structure in only three modes,
permitting a fully analytical treatment. By focusing on
the gauge structure underlying the phase symmetry, we
propose a unified approach to 
characterizing both the insensitivity of the turbulent
energy cascade to symmetry breaking and its crucial role
in linear stability.

\textit{Shell model of turbulence.}---
The shell model possesses a phase symmetry analogous to the Galilean invariance of the Navier--Stokes equations;
this structural similarity motivates us to examine how symmetry breaking by external forcing governs subcritical transition in the shell model.
We consider the Gledzer-Ohkitani-Yamada (GOY) shell model  
for $\bm{X}=(X_1,X_2,\cdots,X_N) \in \mathbb{C}^N$,
where $X_n$ represents a variable at wavenumber $k_n=k_02^{n}$,
driven by the external force $\bm{f}=(f_1,\cdots,f_n)\in \mathbb{C}^N$ with $N=25$ for numerical demonstration.
In addition to the standard form, 
we introduce gauge variables $\bm{U}=(U_1,\cdots,U_N)\in\mathbb{R}^N$ to deal 
with a redundant degree of freedom in the unforced case $\bm{f}=\bm{0}$,
in the following form 
\begin{multline}
  \left(\frac{d}{dt} + \nu k_n^2 - iU_n\right) X_n
  = i\Big(a k_{n+1} X_{n+2}^* X_{n+1}^* \\
  + b k_n X_{n+1}^* X_{n-1}^*
  + c k_{n-1} X_{n-1}^* X_{n-2}^*\Big) + f_n
  \label{eq:GOYshell}
\end{multline}
where the asterisk indicates the complex conjugate.
The nonlinear terms parametrized by $a$, $b$, and $c$ redistribute 
the energy $E(n)\equiv|X_n|^2/2$ and conserve 
the total energy $K\equiv\sum_{n=1}^{N} E(n)$
when $a+b+c=0$.

\textit{How symmetry breaking affects dynamics.}---
We now focus on the phase symmetry and the corresponding invariance properties of the equations.
Because the state variables are complex, phase transformations generate a $\mathrm{U}(1)^N$ symmetry that leaves the energy unchanged. 
In the unforced case, however, the equations of motion are
invariant only under the reduced symmetry
$\mathrm{U}(1) \times \mathrm{U}(1)$: the full $\mathrm{U}(1)^N$ phase rotation
$X_n \to X_n \exp(i\theta_n)$ leaves Eq.~(\ref{eq:GOYshell})
unchanged only when
$\theta_{n+2} + \theta_{n+1} + \theta_n = 0$ (mod $2\pi$)
for all~$n$, which restricts the $N$ independent phases to
two free parameters $\theta_1$ and $\theta_2$.
Under this symmetry, a gauge transformation
$(X_n(t), U_n) \to (e^{iV_n t} X_n(t),\, U_n + V_n)$
with $V_n + V_{n+1} + V_{n+2} = 0$ maps solutions of
Eq.~\eqref{eq:GOYshell} onto each other. Thus, any system
with $U_n + U_{n+1} + U_{n+2} = 0$ is gauge-equivalent to
the reference system $U = 0$, analogous to Galilean frame
equivalence in the Navier--Stokes equations.
Breaking Galilean invariance, for example by fixing a
boundary condition that selects a preferred frame, is
analogous to introducing nonzero~$U$.

To consider the symmetry breaking by the external forcing precisely,
we limit our interest to the case in which the
external forcing acts only on a single mode
$n = N_f$, 
so that $k_{N_f} = k_0 2^{N_f} = 1$.
A laminar solution $\bar{X}_n = \delta_{nN_f}$
is sustained by
$f_n=(\nu k_n^2-iU_n)\delta_{nN_f}$,
where $\delta_{nm}$ is the Kronecker delta.
In the presence of the external force, the phase symmetry is partially broken from $\mathrm{U}(1)\times \mathrm{U}(1)$ to $\mathrm{U}(1)$,
i.e., the phase symmetry for $\theta_{N_f+1+3m}+\theta_{N_f+2+3m}=0$ $(m\in \mathbb{Z})$ remains intact.
Therefore, the shell model with $U_{N_f+1+3m}+U_{N_f+2+3m}=-U_{N_f+3m}$ ($m\in\mathbb{Z}$) 
is equivalent to the reference system in the unforced case but has a different solution with a nonzero forcing strength.
We refer to a system satisfying the gauge
condition $U_n+U_{n+1}+U_{n+2} = 0$ as gauge-equivalent,
while 
a system violating this condition is termed gauge-inequivalent.
Hereafter, $U$ parametrizes the gauge-equivalent case through
$U_{N_f+1+3m} = U_{N_f+2+3m} = -U$ ($m \in \mathbb{Z}$) without loss of generality.

The energy flux through the $n$th shell and its balance
are given by
\begin{equation}
  \Pi(n) \equiv -\operatorname{Im}
  [X_n X_{n+1}(ak_{n+1}X_{n+2} - ck_n X_{n-1})],
  \label{eq:pi}
\end{equation}
\begin{equation}
  \frac{dS(n)}{dt} = \Pi(n) + I(n) + D(n),
  \label{eq:flux}
\end{equation}
where $S(n) \equiv \sum_{m=1}^{n} E(m)$,
$I(n) \equiv \mathrm{Re}\sum_{m=1}^{n} X_m^* f_m$, and
$D(n) \equiv -\nu \sum_{m=1}^{n} k_m^2 |X_m|^2$.
Since $\Pi(n)$ and $D(n)$ are independent of~$U_n$ 
in gauge-equivalent system
while
only $I(n)$ depends on it, 
external phase-symmetry breaking does not alter
the energy flux or dissipation in the inertial range.

\textit{Vanishing linear instability via symmetry breaking}---
The linear stability for $\overline{X_{n}}$ involves only five complex variables
for $n=N_f-2,N_f-1,N_f,N_f+1$ and $N_f+2$ 
because nonlinear terms in Eq.~\eqref{eq:GOYshell} couple to two adjacent variables.
Substituting the ansatz $\bm{X}(t) = \exp(\lambda t)\bm{X}(0) + \overline{\bm{X}}$ into the linearized form of Eqs.~\eqref{eq:GOYshell} around $ \overline{\bm{X}}$
yields that
the eigenvalue problem for $\lambda$ decomposes into
$M_{\mathrm{shell}}\Psi_1 = \lambda\Psi_1$ and
$M_{\mathrm{shell}}^*\Psi_2 = \lambda\Psi_2$,
with $\Psi_1 = (X_{N_f+2},\, X_{N_f+1}^*,\,
X_{N_f-1},\, X_{N_f-2}^*)$ and its conjugate counterpart
$\Psi_2$, where
\begin{equation}
M_{\mathrm{shell}} \equiv
\begin{pmatrix}
  \mu_{+2}  & ick_{N_f\!+\!1} & 0 & 0 \\
  -ib^*k_{N_f\!+\!1} & \mu_{+1}^* & -ic^*k_{N_f} & 0 \\
  0 & iak_{N_f} & \mu_{-1} & ibk_{N_f\!-\!1} \\
  0 & 0 & -ia^*k_{N_f\!-\!1} & \mu_{-2}^*
\end{pmatrix}
\label{eq:Mshell}
\end{equation}
with $\mu_{\pm j} \equiv -\nu k_{N_f \pm j}^2 + iU_{N_f \pm j}$.
The remaining eigenvalues
$\lambda = -\nu k_{N_f}^2 \pm iU_{N_f}$ have negative
real parts, representing exponential decay of
infinitesimal disturbances.

The asymptotic form of the eigenvalue for large $U$ is
obtained by perturbation theory in $|U|^{-1} \ll 1$.
At leading order, $iM_{\mathrm{shell}}$ is diagonal and
Hermitian, so only the diagonal elements contribute at
the next order, yielding
\begin{equation}
  \lambda = -\nu k_{N_f \pm m}^2\pm iU + O(|U|^{-1})
  \quad (m = 0,1,2),
  \label{eq:asym}
\end{equation}
whose real part is always negative. 
This asymptotic form
resembles the dispersion relation of pseudo-Nambu--Goldstone
(pNG) modes: whereas exact symmetry breaking produces
massless Goldstone modes, explicit breaking by an
external field gives them a finite frequency proportional
to the breaking strength~\cite{delacretaz2022damping}. 
Here, the
external forcing breaks the $\mathrm{U}(1)$ phase symmetry, and
the eigenvalues acquire an imaginary part proportional
to~$U$ while their real parts remain governed by viscous
damping alone, independent of the nonlinear coupling
strengths. This constitutes a testable prediction for
stability analysis of the Navier--Stokes
equations with controlled Galilean-symmetry breaking.
 
\begin{figure}
  \centering
  \begin{overpic}[width=0.49\columnwidth]{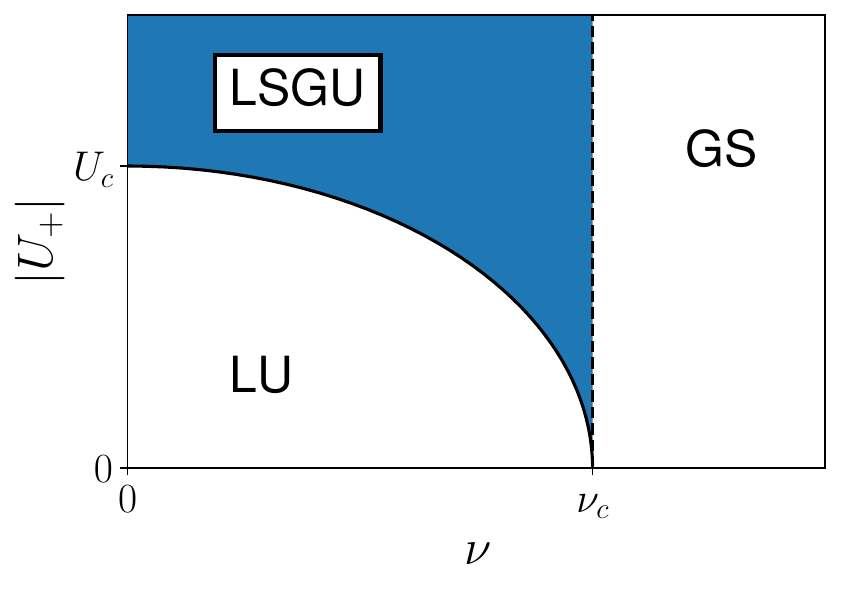}
    \put(2,75){(a)}
  \end{overpic}
  \begin{overpic}[width=0.49\columnwidth]{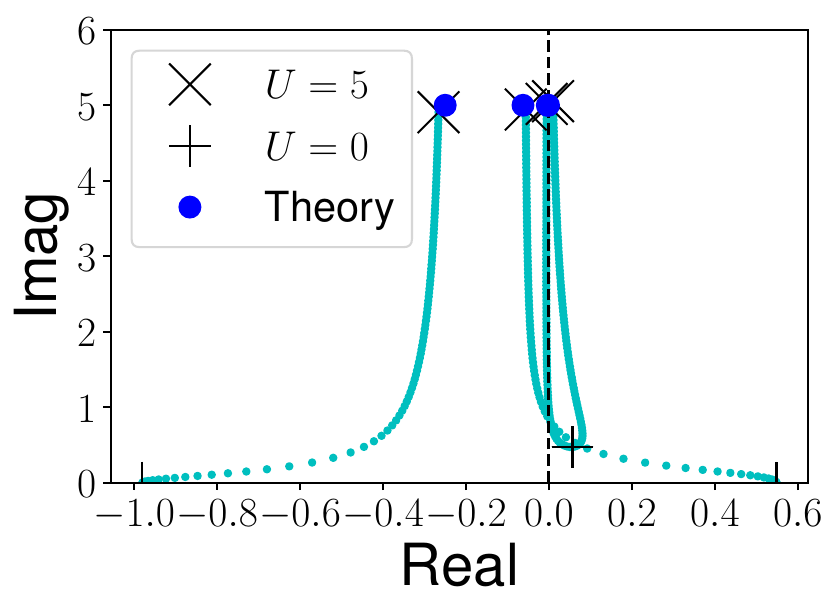}
    \put(2,75){(b)}
  \end{overpic}
  \caption{
    (a)~Neutral stability curve [Eq.~\eqref{eq:oval}] for
  the ST model in the $(\nu,\, |U_+|)$ plane.
  LU: linearly unstable;
  GS: globally stable (bounded by the energy stability
  threshold $\nu_e$);
  LSGU: linearly stable but globally unstable.
  (b)~Eigenvalues of the linear stability problem for the
  shell model at $U = 0$ (dashed) and $U = 5$ (crosses),
  compared with the asymptotic form Eq.~\eqref{eq:asym}
  (circles).%
  \label{fig:oval}
  }
\end{figure}

Figure~\ref{fig:oval}(b) confirms this suppression: at
$U = 0$ the eigenvalues can have positive real parts, but
with increasing~$U$ they approach
Eq.~(\ref{eq:asym}) and the linear instability vanishes
once $|U|$ exceeds a critical value, even when $\nu$ is
small enough for the reference system to be unstable.

\textit{Single triad model and analytical linear stability.}
To further demonstrate the impact of external phase symmetry breaking, 
we introduce another model that
 simplifies the nonlinear interaction in the GOY shell model.
We consider only three variables $(X_1, X_2, X_3)$
\begin{align}
  \frac{d}{dt}X_1&=(-\nu k_1^2+iU_1)X_1+i\tilde{a}X_2^{\ast}X_3^{\ast}+f_1,\label{eq:STM1} \\
  \frac{d}{dt}X_2&=(-\nu k_2^2+iU_2)X_2+i\tilde{b} X_3^{\ast}X_1^{\ast}+f_2,\label{eq:STM2} \\
  \frac{d}{dt}X_3&=(-\nu k_3^2+iU_3)X_3+i\tilde{c}X_1^{\ast}X_2^{\ast}+f_3, \label{eq:STM3}
\end{align}
with time evolution [Eqs.~\eqref{eq:STM1}--\eqref{eq:STM3}].
We refer to this as the single-triad (ST) model.
The coupling coefficients $\tilde{a}$, $\tilde{b}$,
$\tilde{c}$ play the same role as $a$, $b$, $c$ in the
GOY model; $K = \sum_{n=1}^{3}|X_n|^2/2$ is conserved
when $\tilde{a} + \tilde{b} + \tilde{c} = 0$.
As in the shell model, the unforced ST model is invariant
under $X_n \to X_n e^{i\theta_n}$ with
$\theta_1 + \theta_2 + \theta_3 = 0$, and systems with
$U_1 + U_2 + U_3 = 0$ are gauge-equivalent to the
reference system.

The laminar solution $\bar{X} = (\bar{X}_1, 0, 0)$ has
two trivial eigenvalues $\lambda = -\nu k_1^2 \pm iU_1$
and nontrivial eigenvalues determined by
$M_{\mathrm{ST}}\Phi_1 = \lambda\Phi_1$ and
$M_{\mathrm{ST}}^*\Phi_2 = \lambda\Phi_2$ for
$\Phi_1 = (X_2,\, X_3^*)$, with
\begin{equation}
  M_{\mathrm{ST}} =
  \begin{pmatrix}
    -\nu k_2^2 + iU_2 & ib\bar{X}_1^* \\
    -ic\bar{X}_1 & -\nu k_3^2 - iU_3
  \end{pmatrix}.
  \label{eq:MST}
\end{equation}
The eigenvalues are
$\lambda = -\nu(k_2^2 + k_3^2)/2 + i(U_2 - U_3)/2
\pm \sqrt{\Delta}$, where
$\Delta = [\nu(k_3^2 - k_2^2)/2 + iU_+]^2
+ bc|\bar{X}_1|^2$ with
$U_+ \equiv (U_2 + U_3)/2$.
The condition $\mathrm{Re}(\lambda) = 0$ yields the
elliptic neutral curve
\begin{equation}
  \frac{\nu^2}{\nu_c^2} + \frac{U_+^2}{U_c^2} = 1,
  \label{eq:oval}
\end{equation}
with $\nu_c \equiv |\bar{X}_1|\sqrt{bc}/|k_2 k_3|$
and $U_c \equiv (k_2^2 + k_3^2)\nu_c/2$
[FIG.~\ref{fig:oval}(a)].

Equation~\eqref{eq:oval} shows that the laminar state
is linearly unstable for $\nu < \nu_c$ at $U_+ = 0$, that
the instability threshold decreases monotonically with
$|U_+|$, and that linear stability holds for all~$\nu$
once $|U_+| > U_c$ [FIG.~\ref{fig:oval}(a)]. This opens
a linearly stable but globally unstable (LSGU) 
regime, the defining feature of subcritical transition,
confirmed
numerically below. Since the asymptotic form
Eq.~\eqref{eq:asym} depends only on $|U|$ and not on
the coupling coefficients, the suppression of linear
instability is a structural consequence of the
anti-Hermitian component induced by phase-symmetry
breaking. The ST model thus serves as a minimal framework
that isolates this mechanism analytically.

\textit{Nonlinear states and developed turbulence.}---
We now confirm that nonlinear states persist in the
linearly stable regime for both the ST and shell models
with $(a, b, c) = (\tilde{a}, \tilde{b}, \tilde{c}) = (1, -0.5, -0.5)$.
For the ST model with $(k_1, k_2, k_3) = (1, 2, 3)$,
$(U_1, U_2, U_3) = (-2U,\, U,\, U)$,
and $\nu = 10^{-3} < \nu_c \simeq 0.083$,
we vary the initial disturbance amplitude~$h$ in
$X_n(0) = \bar{X}_n + 2h\,e^{i\pi/4}$ and monitor the
disturbance energy $E_d \equiv \sum_n |X_n - \bar{X}_n|^2/2$.
At $U = 0.6 > U_c \simeq 0.54$, small disturbances
($h = 0.1$, $0.44$) decay to the laminar state,
while larger ones ($h = 0.71$, $1$) sustain a nonlinear
state [FIG.~2(a)], demonstrating the subcritical character
of the transition.
The shell model exhibits the same qualitative behavior.
With $(U_{N_f+3m},\, U_{N_f+1+3m},\, U_{N_f+2+3m}) = (U,\, U,\, -2U)$,
$\nu = 10^{-6}$, $\bar{X}_n = \delta_{n5}$,
and a linear drag $\nu_0 = 10^{-3}$ for $n \le 7$,
the eigenvalue problem for $M_{\mathrm{shell}}$ gives
$U_c \simeq 0.765$ [FIG.~1(b)].
At $U = 0.8 > U_c$, a small disturbance ($h = 0.001$)
decays with slow periodic modulation reflecting the small
viscosity, whereas larger disturbances
($h = 0.01$, $0.1$) develop into chaotic motion
[FIG.~2(b)].
Moreover, for both reference systems, $E_d$ relaxes toward a specific value independent of $h$.
Although the ST model does not quantitatively reproduce
the shell model dynamics, both share the monotonic
suppression of linear instability with increasing $|U|$
[FIG.~1(b)], supporting the interpretation that the
mechanism identified analytically in the ST model
applies to
the full shell model.

\begin{figure}[t]
  \centering
  \begin{overpic}[width=0.49\columnwidth]{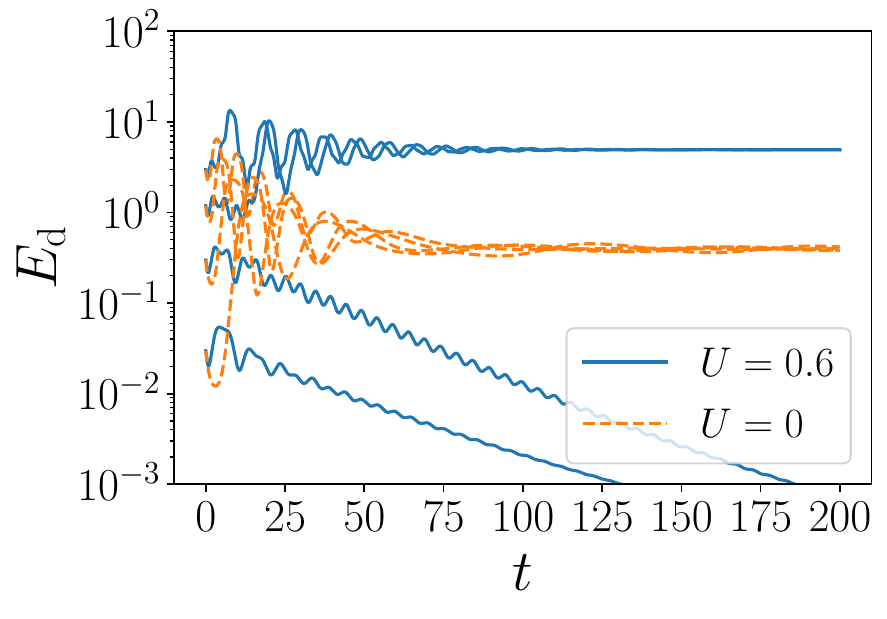}
    \put(2,70){(a)}
  \end{overpic}
  \begin{overpic}[width=0.49\columnwidth]{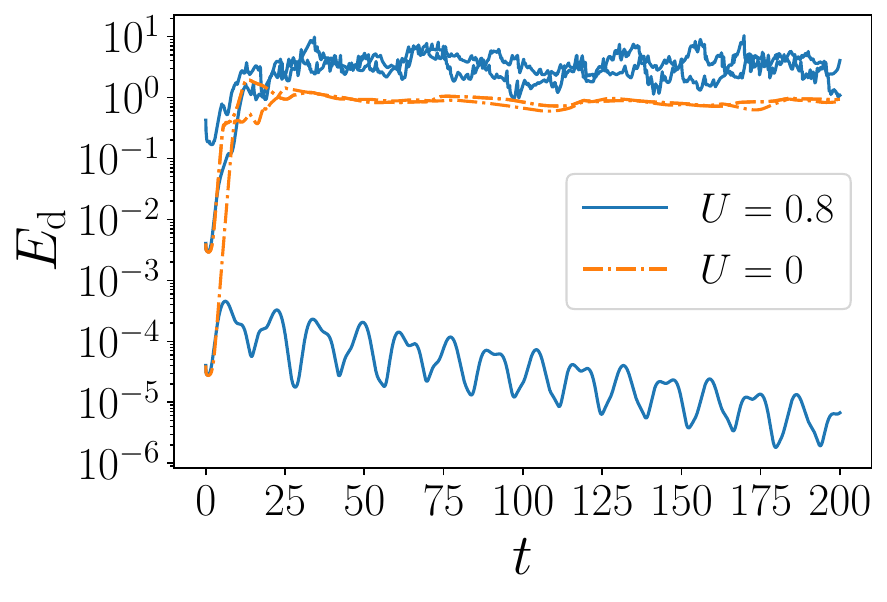}
    \put(2,70){(b)}
  \end{overpic}
  \caption{\label{fig:E1} 
    Disturbance energy $E_d(t)$ for various initial
  amplitudes~$h$.
  (a)~ST model with $\nu = 10^{-3}$:
  solid lines, $U = 0.6 > U_c$;
  dashed lines, reference system ($U = 0$).
  (b)~Shell model with $\nu = 10^{-6}$:
  solid lines, $U = 0.8 > U_c$;
  dashed lines, reference system ($U = 0$).
  In both cases, small disturbances decay 
  in the gauge-equivalent system while large
  disturbances
  sustain nonlinear states.%
  }
\end{figure}

\begin{figure}[t]
  \centering
  \begin{overpic}[width=\columnwidth]{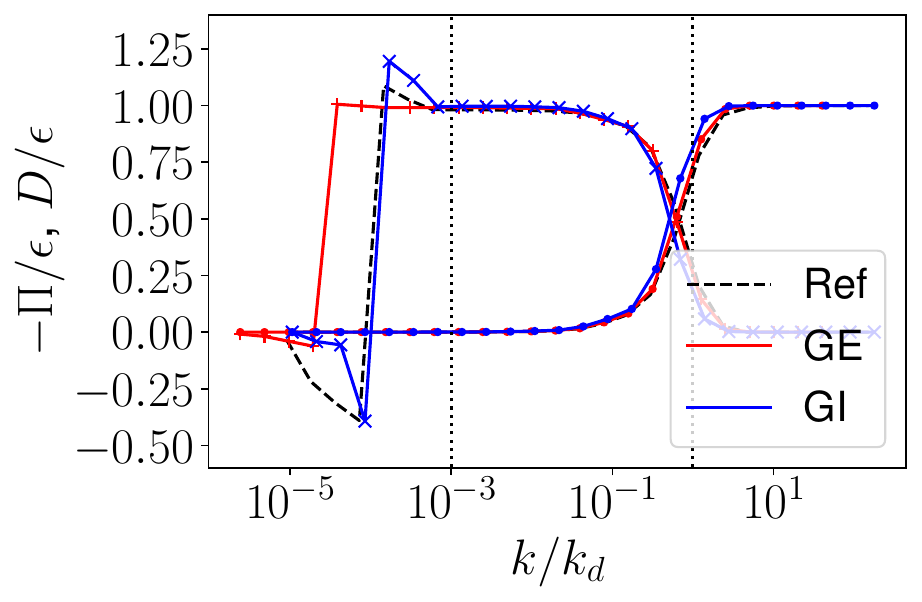}
    \put(2,60){(a)}
  \end{overpic}\\
  \begin{overpic}[width=\columnwidth]{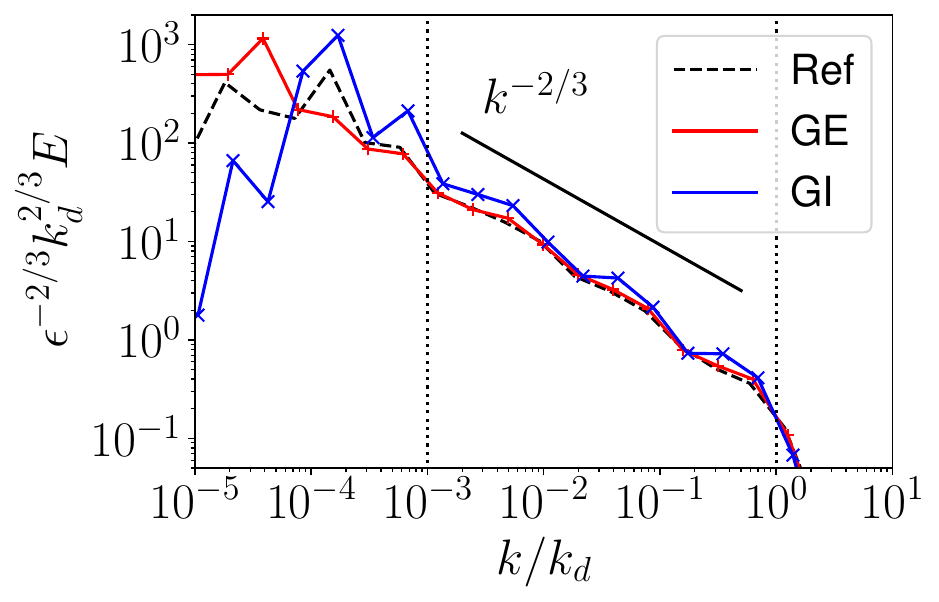}
    \put(2,60){(b)}
  \end{overpic}
  \caption{\label{fig:E4}
  (a)~Energy flux $-\Pi(n)/\epsilon$ and dissipation
  $D(n)/\epsilon$ 
  for the shell
  model.
  (b)~Energy spectrum
  $\epsilon^{-2/3}k_d^{2/3}E$.
  Plus markers: gauge-equivalent system~(GE),
  $(U_{N_f+3m},\,U_{N_f+1+3m},\,U_{N_f+2+3m}) = (U,\,U,\,-2U)$;
  crosses: gauge-inequivalent system~(GI),
  $(U_{N_f+3m},\,U_{N_f+1+3m},\,U_{N_f+2+3m}) = (U,\,U,\,0)$.
  Dashed lines: reference system ($U=0$).
  Vertical dotted lines mark $k/k_d = 10^{-3}$ and~$1$.%
  }
\end{figure}

In the steady state, the energy flux statistically balances dissipation
as $\epsilon = -\Pi(n)$ at intermediate scales, yielding
$E(n) \propto k_n^{-2/3}$; the dissipation wave number is
$k_d \equiv \epsilon^{1/4}\nu^{-3/4}$.
To verify that the essential features of the turbulent
state are preserved under external phase-symmetry
breaking, we examine
the energy flux
$\Pi(n)$ and dissipation $D(n)$ normalized by
$\epsilon = D(N)$. A constant energy flux appears at
intermediate scales $10^{-3} < k/k_d < 10^{-1}$ for both
the reference and gauge-equivalent systems
[FIG.~\ref{fig:E4}(a)],
and the normalized
energy spectrum [FIG.~\ref{fig:E4}(b)] agrees well with the reference
system, including the period-3 oscillation characteristic
of the GOY model, even at $U = 0.8$.
In contrast, under gauge-inequivalent system with
$U_n + U_{n+1} + U_{n+2} \neq 0$, the energy flux
Eq.~\eqref{eq:pi} acquires an additional phase factor
$\exp(i(U_n + U_{n+1} + U_{n+2}))$, 
altering the functional form of $\Pi(n)$. Since
the energy flux remains constant even with the
modified $\Pi$ [FIG.~\ref{fig:E4}(a)], the energy spectrum
must adjust accordingly, resulting in deviation from
the reference system at intermediate scales
[FIG.~\ref{fig:E4}(b)], while the linear stability remains
identical to the gauge-equivalent case.
This confirms that
the spectral equivalence with the reference system is
governed by the phase-symmetry condition rather than by
the value of~$U$ alone.

\textit{Summary and conclusion.}---
In this Letter, we investigated the linear stability of the GOY shell model
 incorporating phase symmetry breaking due to external forcing, 
motivated by the Galilean invariance of the Navier--Stokes equations.
We introduced a gauge variable to handle the redundant degrees of freedom, 
and the perturbation theory confirms the suppression of linear instability.
To further demonstrate this stability problem, we have introduced the single triad model 
to prove the vanishing linear instability analytically.
Furthermore, in the gauge-equivalent system, 
the enhanced linear stability is shown not to alter the
inter-scale energy flux or the energy spectrum in the
shell model.

The linear stabilization of laminar solutions is understood mathematically 
through the emergence of an anti-Hermitian component in the eigenvalue problem of the linearized equations. 
In this Letter, we demonstrated that such an
anti-Hermitian component naturally 
arises from the phase-symmetry breaking in the nonlinear terms of the governing equations.
This symmetry corresponds to invariance under Galilean transformations, and similar conclusions can be expected not only for the Navier--Stokes equations but also for general Galilean invariant equations.

In the present study, we have considered a shell model in which interactions are local in length scale. 
The insensitivity of the present results to the detailed form of the nonlinear interactions, as indicated by the perturbative analysis, reflects the fact that 
the suppression of linear instability originates from the symmetry structure of the linearized operator rather than from the specific coupling coefficients. 
This robustness supports the use of the ST model as a minimal but representative framework for understanding subcritical transition via phase symmetry breaking in shell models.
Indeed, in Navier--Stokes systems, conditions that fix the
Galilean frame of the laminar state are known to control
the nature of the transition: in two-dimensional
Kolmogorov flow, the imposed mean flow rate determines
whether the transition is supercritical or
subcritical~\cite{hiruta_2020,hiruta2022}, while in
Taylor--Couette--Poiseuille flow, an axial pressure
gradient switches between the two transition
types~\cite{matsukawa2025switching}. In the framework developed here, such
frame-fixing conditions correspond to gauge-equivalent
systems, thereby
modifying the linear stability of the laminar state.
Whether this correspondence can be made quantitative, and
whether the suppression of linear instability in the
gauge-equivalent case, while preserving the essential
features of the turbulent state, 
extends to the Navier--Stokes equations, 
remains an important
open question.

\begin{acknowledgments}
Y.H. acknowledges the Japan Society for the Promotion of Science (JSPS) KAKENHI (Grant No. 21H05309)
and JST CREST Japan (Grant Number JPMJCR25Q1).
\end{acknowledgments}

\bibliography{bib}

\end{document}